\documentclass[sigconf,screen,authorversion,nonacm]{acmart}
\AtBeginDocument{}

\usepackage{xspace}
\usepackage{listings}

\usepackage{amssymb}
\usepackage{pifont}

\usepackage{subcaption}
\usepackage{float}

\newcommand{\cmark}{\ding{51}}\newcommand{\xmark}{\ding{55}}\usepackage{multirow}
\usepackage{multicol}
\usepackage{fancyvrb}
\usepackage[linesnumbered, ruled, vlined,algo2e, procnumbered, algonl]{algorithm2e}

\definecolor{AF00DB}{rgb}{0.69, 0, 0.86}
\definecolor{098658}{rgb}{0.04, 0.53, 0.35}
\definecolor{001080}{rgb}{0, 0.06, 0.5}
\definecolor{795E26}{rgb}{0.47, 0.37, 0.15}
\definecolor{0000FF}{rgb}{0, 0, 1}
\definecolor{AF00DB}{rgb}{0.69, 0, 0.86}
\definecolor{008000}{rgb}{0, 0.5, 0}
\definecolor{A31515}{rgb}{0.64, 0.08, 0.08}
\definecolor{267F99}{rgb}{0.15, 0.5, 0.6}
\definecolor{EE0000}{rgb}{0.93, 0, 0}

\definecolor{comment}{rgb}{0,0.6,0}
\definecolor{keyword}{rgb}{0,0.33,0.58}
\definecolor{mygray}{rgb}{0.5,0.5,0.5}
\definecolor{darkgray}{rgb}{0.2,0.2,0.2}
\definecolor{string}{rgb}{0.72,0.33,0}
\lstdefinestyle{Mystyle}{
	backgroundcolor=\color{white},   basicstyle=\footnotesize\ttfamily,        breakatwhitespace=false,         breaklines=true,                 captionpos=b,                    commentstyle=\color{comment},    deletekeywords={...},            extendedchars=true,              frame=lines,                    keepspaces=true,                 numbers=left,                    numbersep=4pt,                   numberstyle=\tiny\color{darkgray}, rulecolor=\color{black},         showspaces=false,                showstringspaces=false,          showtabs=false,                  tabsize=2,                       xleftmargin=8pt,
}
\setcitestyle{numbers,sort&compress}

\usepackage{url}

\definecolor{346fce}{rgb}{0.20, 0.44, 0.81}

\definecolor{fa76b3}{rgb}{0.98, 0.46, 0.70}

\providetoggle{beginFlag}
\settoggle{beginFlag}{true}

\SetAlgoSkip{CustomSkipAlgoConfig}

\newcommand{\mycomment}[1]{}
\newcommand{\FIGREF}{Figure\xspace}

\renewcommand\appendix{\par
	\setcounter{section}{0}
	\setcounter{subsection}{0}
	\setcounter{figure}{0}
	\setcounter{table}{0}
	\setcounter{lstlisting}{0}
	\renewcommand\thesection{Appendix \Alph{section}}
	\renewcommand\thefigure{\Alph{section}\arabic{figure}}
	\renewcommand\thetable{\Alph{section}\arabic{table}}
	\renewcommand\thelstlisting{\Alph{section}\arabic{lstlisting}}
	\setcounter{algocf}{0}
	\renewcommand{\thealgocf}{\Alph{section}\arabic{algocf}}
	\renewcommand*{\theHsection}{\thesection}
	\renewcommand*{\theHfigure}{\thefigure}

}
\SetArgSty{textnormal}
\SetKwInOut{Input}{Input}
\SetKwInOut{Output}{Output}
\SetKw{Continue}{continue}
\SetKw{Break}{break}
\SetKw{IfInline}{if}
\SetKw{Goto}{goto}
\SetKwProg{try}{try}{:}{}
\SetKwProg{catch}{catch}{:}{end}

\SetKwComment{Comment}{/* }{ */}
\SetKwRepeat{Repeat}{repeat}{until}

\SetCommentSty{mycommfont}
\newcommand{\assignment}{\ensuremath{\gets}\xspace}

\newcounter{Chall}

\definecolor{codegreen}{rgb}{0,0.6,0}
\definecolor{codegray}{rgb}{0.5,0.5,0.5}
\definecolor{codepurple}{rgb}{0.58,0,0.82}
\definecolor{backcolour}{rgb}{0.95,0.95,0.92}
\lstdefinestyle{promptstyle}{
	backgroundcolor=\color{backcolour},
	commentstyle=\color{codegreen},
	keywordstyle=\color{magenta},
stringstyle=\color{codepurple},
	basicstyle=\ttfamily\scriptsize,
	breakatwhitespace=true,
	breaklines=true,
	captionpos=b,
	keepspaces=false,
	numbers=left,                    numbersep=4pt,                   numberstyle=\tiny\color{darkgray}, rulecolor=\color{black},         showspaces=false,
	showstringspaces=false,
	showtabs=false,
	tabsize=2,	
	xleftmargin=8pt,
	xrightmargin=5pt,
morekeywords={rule,when,Item,received,then,end,Input,Output,Role,Contexts,Task, Policy, Text,Tutorial,Examples,Instruction,Refinement,Constraints,Context}
}
\definecolor{myfunctioncolor}{rgb}{0.51,0.09,0.39} \lstdefinestyle{customc}{
	belowcaptionskip=0.5\baselineskip,
	breaklines=true,
tabsize=2,
language=C,
	showstringspaces=false,
	basicstyle=\scriptsize\ttfamily,
keywordstyle=\color{teal!75!black}\itshape,
	commentstyle=\itshape\color{orange!65!black}\mdseries,
	identifierstyle=\color{blue!60!black},
	stringstyle=\color{green!60!yellow!50!black},
	numbers=left,
	numbersep=4pt,                   numberstyle=\tiny\color{darkgray}, rulecolor=\color{black},         showspaces=false,                showstringspaces=false,          showtabs=false,                  tabsize=2,                       xleftmargin=8pt,	
morekeywords=[2]{main, printf, scanf, socket, send,fuzz, perror, memset, read, setsockopt,bind,listen,accept,connect}, keywordstyle=[2]{\color{myfunctioncolor}},
}
\copyrightyear{2026}
\acmYear{2026}
\begin{document}
	

	\title{Evaluating the Effectiveness of LLMs in Aiding Compliance Testing of PKCS\#1-v1.5}
	
	\author{Polina Kozyreva}
	\orcid{0009-0009-8575-5472}
	\affiliation{
		\institution{Syracuse University}
		\city{Syracuse}
		\state{NY}
		\country{USA}}
	\email{pkozyrev@syr.edu}
	\author{Endadul Hoque}
	\orcid{0000-0002-6682-9618}
	\affiliation{
		\institution{Syracuse University}
		\city{Syracuse}
		\state{NY}
		\country{USA}}
	\email{enhoque@syr.edu}

\begin{abstract}
Testing implementations of binary protocols for specification compliance requires inputs that satisfy both structural and semantic constraints. Purely random generation and primitive mutations are often insufficient for exploring semantically meaningful behaviors in protocols that rely on Type-Length-Value (TLV) encoding, yet domain-specific compliance testing tools require deep protocol expertise and significant manual effort to construct. This work investigates whether grammar-level mutation combined with LLM-based code synthesis can serve as a viable, more generalizable approach to specification compliance testing. 

We evaluate the approach on PKCS\#1~v1.5 signature verification--a widely deployed TLV-encoded standard with a formally verified testing oracle (Morpheus)--across 48 cryptographic library implementations. We reproduced 10 of 13 non-trivial specification violation categories previously identified by Morpheus, including all 5 signature forgery categories, and discovered 1 previously unreported discrepancy. We found that LLM hallucination--occurring in 82.5\% of generated scripts--is the primary factor limiting effectiveness, not the mutation strategies. We identify five distinct hallucination types and show that their distribution varies systematically across mutation categories: structural mutations are implemented with 13.3\% fidelity while constraint mutations achieve 30.3\% correctness but suffer the highest rate of mutations being fully ignored (8.1\%). These findings reveal a striking gap between operational reliability (99.8\%) and semantic fidelity (17.5\%), providing actionable guidance on when LLM-based code synthesis can be trusted in specification-driven testing pipelines
\end{abstract} \maketitle

\vspace*{-1.\baselineskip}
\section{Introduction}
\label{sec:intro}
Protocol implementations must conform to their specifications--typically expressed as RFCs--in order to guarantee interoperability and security. Deviations from the specification, even when they do not cause crashes, can lead to exploitable vulnerabilities. A prominent example is the Bleichenbacher low-exponent RSA signature forgery~\cite{bleichenbacher1998chosen}, which exploits lax parsing of PKCS\#1~v1.5 encoded structures: an implementation that accepts malformed padding or ignores trailing bytes can be tricked into accepting a forged signature.

Detecting such specification-compliance violations is challenging because it requires test inputs that satisfy both the structural constraints of the protocol format (e.g., Type-Length-Value encoding, hierarchical nesting) and the semantic constraints imposed by the specification (e.g., length-value consistency, padding rules). Purely random input generation is insufficient: the probability of randomly producing a byte sequence that passes initial format validation and exercises deeper parsing logic is negligible for binary protocols with complex structure.

Existing approaches to this problem fall into two classes, namely, general and domain-specific. At one end, domain-specific tools such as Morpheus~\cite{yahyazadeh2021morpheus} construct formal models of a specific protocol and exhaustively generate test cases from the model. Morpheus tested 45 PKCS\#1~v1.5 implementations and uncovered signature forgery vulnerabilities in several of them, but its combinatorial model required deep protocol expertise and significant manual effort to construct, limiting generalizability. At the other end, general-purpose fuzzers such as AFLNET~\cite{pham2020aflnet} and coverage-guided tools like AFL++~\cite{fioraldi2020afl++} can be applied to any target but lack awareness of protocol semantics, making them unable to systematically generate inputs that test specification compliance.

Recent work has begun to explore new directions that lie in-between these two classes. FANDANGO~\cite{zamudio2025fandango} uses a genetic algorithm to evolve inputs that satisfy grammar-and-constraint specifications. ChatAFL~\cite{meng2024chatafl} leverages LLMs to extract protocol grammars and enrich fuzzing seeds. Gmutator~\cite{bendrissou2025grammar} mutates grammar production rules themselves, but operates only on syntactic structure and explicitly identifies semantic constraint mutation as future work. However, none of these approaches mutates both structural rules and semantic constraints at the grammar level: FANDANGO and ChatAFL mutate inputs within a fixed grammar, while Gmutator mutates grammar structure but leaves constraints untouched.

\textbf{This work investigates whether grammar-level mutation, targeting both structural rules and semantic constraints, combined with LLM-based code synthesis, is a viable approach to specification compliance testing of binary protocols.} As means to answer this question we have designed the simplest pipeline that mutate both rules and constraints of the original grammar and use an LLM to translate each mutated variant into an executable input generator. The test cases produced from the input generators would exercise both valid and invalid regions of the input space, enabling detection of specification-level inconsistencies across implementations. We evaluate on PKCS\#1~v1.5, a widely deployed cryptographic standard whose nested TLV structure and strict semantic constraints make it a representative and challenging target, and for which Morpheus provides a formally verified oracle enabling rigorous comparison. Our findings demonstrate that the approach is viable--reproducing the majority of known violations and uncovering a new one--but reveal that LLM semantic fidelity, not the mutation strategy, is the primary bottleneck. This finding has implications beyond our specific study: any approach that uses LLMs to generate code from formal specifications must contend with the gap between operational reliability and semantic correctness.

Our contributions are:
\begin{itemize}
	\item An empirical investigation of whether grammar-level mutation combined with LLM-based code synthesis can serve as a viable approach to specification compliance testing of binary protocols, instantiated through 13 mutation operators across structural and constraint dimensions.
	\item Evidence that the approach reproduces 10 of 13 (76.9\%) non-trivial violation categories identified by Morpheus--including all 5 signature forgery categories--and uncovers 1 previously unreported discrepancy, across 48 PKCS\#1~v1.5 implementations.
	\item A taxonomy and quantitative characterization of LLM hallucination in specification-constrained code generation, identifying five hallucination types across all 1,500 generated scripts and revealing that hallucination patterns differ systematically between structural and constraint mutations--with a striking gap between operational reliability (99.8\%) and semantic fidelity (17.5\%).
\end{itemize}

\vspace*{-1.\baselineskip}
\section{Background}
\label{sec:background}
This section provides background on the PKCS\#1~v1.5 signature scheme, the ASN.1 encoding rules that govern its binary representation, and the classes of specification violations that arise in practice.

\vspace*{-1.\baselineskip}

\subsection{PKCS\#1 v1.5 Signature Scheme}

PKCS\#1 v1.5, specified in RFC 8017 \cite{kaliski2016rfc}, defines the RSA signature scheme. To sign a message $m$, the signer first computes the hash $H(m)$ using a designated hash algorithm (e.g., SHA-256) and then constructs an \emph{encoded message} (EM) of the following form:

\begin{center}
	\texttt{EM = 0x00 || 0x01 || PS || 0x00 || T}
\end{center}

\noindent where \texttt{0x00} is the leading byte, \texttt{0x01} is the block type indicating a signature operation, \texttt{PS} is a padding string consisting entirely of \texttt{0xFF} bytes with a minimum length of 8 bytes, \texttt{0x00} is a separator byte, and \texttt{T} is the DigestInfo value--an ASN.1 DER-encoded structure that contains the hash algorithm identifier and the hash value $H(m)$. The total length of EM must equal the RSA key size in bytes (e.g., 256 bytes for a 2048-bit key).

The signature $S$ is produced using the formula $S = EM^d$ $mod$ $n$ where $d$ is signer's private exponent and $n$ is public modulus. To verify, the recipient performs the following operation $S^e$ $mod$ $n$, $e$ being signer's public exponent, which must match the initial $EM$.

\subsection{ASN.1 and DER Encoding}

The DigestInfo value \texttt{T} within the encoded message is specified using Abstract Syntax Notation One (ASN.1) and encoded using the Distinguished Encoding Rules (DER), as defined in ITU-T X.690 \cite{ITUT_X690_2021}. DER uses a Type-Length-Value (TLV) encoding scheme in which each data element consists of three parts:

\begin{itemize}
	\item \textbf{Type}: identifies the data type (e.g., \texttt{0x30} for SEQUENCE, \texttt{0x06} for OBJECT IDENTIFIER, \texttt{0x04} for OCTET STRING, \texttt{0x05} for NULL).
	\item \textbf{Length}: specifies the number of bytes in the value field. DER supports both short-form (single byte, for lengths $\leq$~127) and long-form (multi-byte, prefixed with \texttt{0x81}--\texttt{0x84}) length encoding.
	\item \textbf{Value}: the content bytes, whose interpretation depends on the type.
\end{itemize}

The semantic constraints of this encoding are strict: each length field must exactly equal the byte-length of its corresponding value, the OID must match the hash algorithm used, and NULL parameter in the AlgorithmIdentifier may be either present or absent, compliant implementations must accept both forms. These constraints, combined with the outer PKCS\#1 padding structure, create a complex format where implementations can deviate from the specification in numerous ways.

\subsection{Specification Violations in Practice}
The verification of PKCS\#1 v1.5 signatures requires checking multiple structural and semantic properties. Morpheus \cite{yahyazadeh2021morpheus} categorizes specification violations into the following classes:

\begin{itemize}
	\item \textbf{Signature Forgery (SF)}: The implementation accepts an input whose structure deviates from the specification in a way that could enable an attacker to forge a valid-looking signature without knowledge of the private key. Examples include accepting inputs with trailing garbage bytes after the DigestInfo, accepting incorrect padding, or accepting modified ASN.1 structures. These are the most security-critical violations.
	
	\item \textbf{Minor Leniency (ML)}: The implementation accepts a technically non-compliant input that does not directly enable signature forgery but represents a deviation from the specification. Examples include accepting long-form length encoding where short-form is required by DER, accepting malformed NULL parameter in the AlgorithmIdentifier, or tolerating minor structural variations within the DigestInfo.
	
	\item \textbf{Buffer Overflow (BO)}: The implementation reads beyond the bounds of the input buffer when processing a malformed structure, potentially leaking memory contents or causing a crash.
	
	\item \textbf{Incompatibility (IN)}: The implementation rejects PKCS\#1 v1.5 encoded messages in which the NULL parameter in the AlgorithmIdentifier is absent, despite RFC 8017 requiring acceptance of both forms.
\end{itemize}

\subsection{LLM-Based Code Generation}
Large language models have demonstrated strong performance on code generation benchmarks, with recent models achieving 49\% on CyberSecEval-3 \cite{potter2025frontier} and showing the ability to produce syntactically correct programs from natural language descriptions. This capability has motivated their use in software testing: SeedMind \cite{shi2024harnessing} uses LLMs to generate input generators from harness code, ChatAFL \cite{meng2024chatafl} uses them to extract protocol grammars, and Fuzz4All \cite{xia2024fuzz4all} uses them to generate test inputs directly.
However, these applications typically provide the LLM with informal or partial specifications. When the task requires strict adherence to a formal specification--such as generating code that produces binary outputs satisfying specific TLV structural constraints and arithmetic relationships between fields--the extent to which LLMs faithfully implement the specification is unknown. Our work addresses this gap: we systematically evaluate how reliably an LLM translates formal grammars with semantic constraints into correct input generators, and characterize the failure type when it does not.

\section{Our Approach}
\label{sec:approach}
\begin{figure}[t]
	\centering
	\vspace{-1\baselineskip}
	\includegraphics[width=\columnwidth]{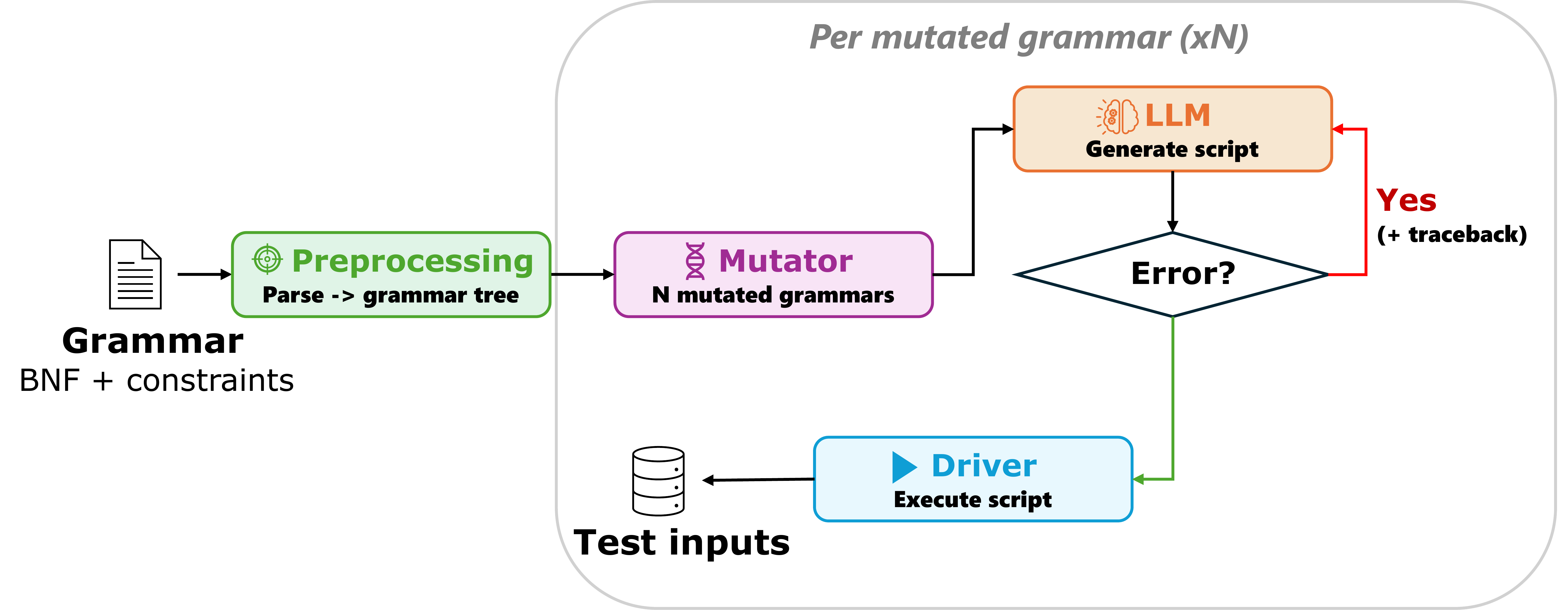}
	\caption{Pipeline architecture. The mutator produces $N$ structurally distinct mutated grammars from the original. Each mutated grammar is sent to the LLM, which generates a standalone Python script (\texttt{gen\_i.py}). The driver executes each script to produce $K$ binary test cases that are stored in a seed bank.}
	\label{fig:pipeline}
	\vspace{-1\baselineskip}
\end{figure}

This section describes the design and implementation that we used to assess the viability of the proposed approach. The pipeline in \FIGREF \ref{fig:pipeline} takes as input a grammar file describing a binary protocol's structure and semantic constraints, and produces a corpus of test cases that can be executed against target implementations to detect specification violations. The pipeline consists of four stages: grammar representation and preprocessing, grammar mutation, LLM-based code synthesis, and test case generation.

\vspace*{-0.5\baselineskip}

\subsection{Grammar Representation}
\label{sec:grammar}

The grammar for PKCS\#1~v1.5 was manually constructed from RFC~8017~\cite{kaliski2016rfc}. The grammar is expressed in Backus-Naur Form (BNF) augmented with semantic constraints, consisting of two parts: production rules and constraint expressions.

Production rules define the hierarchical TLV structure of the encoded message. The grammar contains two kinds of rules. \emph{Structure rules} define composition--for example, the top-level message rule specifies the concatenation of seven components corresponding to the PKCS\#1~v1.5 fields described in Section~\ref{sec:background}:

\smallskip
\noindent\texttt{<message> ::= <leading\_byte> <block\_type>}\\
\null\qquad\texttt{<padding> <separator> <asb\_type>}\\
\null\qquad\texttt{<asb\_length> <asb\_value>;}
\smallskip

\noindent \emph{Terminal rules} assign concrete byte values to nonterminals (e.g., \texttt{<block\_type> ::= "01";} specifies the block type as the byte \texttt{0x01}). Recursive rules define variable-length fields; for example, padding is defined as \texttt{<padding> ::= "ff" <padding> | "";} allowing any number of \texttt{0xFF} bytes.

Semantic constraints are expressed as arithmetic and relational expressions over the nonterminal symbols defined in the production rules. They enforce the protocol's semantic requirements, for example, that TLV length fields must be equal to the byte-length of their corresponding values, that the padding must meet a minimum length, and that the total message length must correspond to a valid RSA key size. As an example, the minimum padding constraint is expressed as \texttt{len(<padding>) >= 16;} where the factor of 16 (rather than 8) arises because the grammar operates on hex-character strings where each byte is represented by one character instead of two. The full grammar contains 20 production rules and 8 constraints.

During preprocessing, the grammar file is parsed into an in-memory \emph{grammar tree} in which nodes correspond to nonterminals or terminals and edges represent composition relationships. Constraints are attached to the relevant nodes. This tree serves as the input to the mutator.

\vspace*{-0.5\baselineskip}

\subsection{Grammar Mutation}
\label{sec:mutations}

The mutator operates on the grammar tree and produces \emph{mutated grammar trees}, each representing a variant of the original specification. We designed 13 mutation operators organized into three categories based on what aspect of the grammar they target and whether they account for TLV structure. The operators are not invented from scratch; rather, they are adapted from mutation strategies employed by existing grammar-based and protocol testing tools--including Gmutator~\cite{bendrissou2025grammar}, Peach~\cite{peach}, and Boofuzz~\cite{boofuzz}--and lifted to operate at the grammar level rather than the input level. The constraint mutation operators (\texttt{remove\_constr}, \texttt{flip\_signs}, \texttt{exchange\_subexpressions}) are new, as no prior tool mutates semantic constraints attached to a grammar.

\vspace*{-0.5\baselineskip}

\subsubsection{Structural Mutations (TLV-Ignorant)}
These nine operators modify the composition or values of grammar components without regard to TLV field relationships:

\begin{itemize}
	\item \texttt{remove\_component}: selects a nonterminal from a structure rule and removes all production rules and constraints that reference it, effectively deleting a protocol field from the specification.
	\item \texttt{reuse\_value}: selects two terminal rules and replaces the value of one with the other's, causing two fields to share the same byte content.
	\item \texttt{repeat\_value}: selects a terminal rule and repeats its value a random number of times, extending the field beyond its specified length.
	\item \texttt{swap\_value}: selects two terminal rules and exchanges their values.
	\item \texttt{swap\_order}: selects a structure rule and transposes two of its components, reordering fields within the message.
	\item \texttt{extend\_tag}: prepends a terminal's value with \texttt{0x1F}, the ASN.1 long-form tag indicator, testing whether implementations handle unexpected tag encodings.
	\item \texttt{long\_length\_form}: prepends a terminal's value with \texttt{0x81}, forcing long-form DER length encoding where short-form would normally be used.
	\item \texttt{byte\_steal}: selects two terminal rules, removes a random number of bytes from the end of one, and appends them to the other, shifting data across field boundaries.
	\item \texttt{add\_garbage}: appends one garbage byte \texttt{0x88} after the structured content, testing whether implementations ignore trailing data.
\end{itemize}

\subsubsection{Structural Mutations (TLV-Aware)}
This operator modifies TLV components while preserving internal consistency:

\begin{itemize}
	\item \texttt{change\_length\_change\_value}: selects a TLV component and increments or decrements its length field and depending on the action chosen either appends a single byte to value field or truncates it by one byte. Unlike the TLV-ignorant operators, this maintains the length-value relationship within the mutated component while creating a mismatch with the surrounding structure.
\end{itemize}

\subsubsection{Constraint Mutations}
These three operators target the semantic constraints rather than the production rules:

\begin{itemize}
	\item \texttt{remove\_constr}: removes a randomly selected constraint from the grammar entirely, eliminating one semantic requirement from the specification.
	\item \texttt{flip\_signs}: selects a constraint and negates all relational operators within it (e.g., $\geq$ becomes $<$, $=$ becomes $\neq$), inverting the semantic requirement.
	\item \texttt{exchange\_subexpressions}: selects two constraints and swaps their sub-expressions, creating novel constraint combinations.
\end{itemize}

An important property of constraint mutations is that they can produce \emph{contradictory grammars}--grammars in which the constraints cannot be simultaneously satisfied. For example, \texttt{flip\_signs} applied to the minimum padding constraint changes \texttt{len(<padding>) >= 16} to \texttt{len(<padding>) < 16}, which may conflict with the total message length constraint that requires padding to fill the remaining space. This property is relevant to the hallucination analysis in Section~\ref{sec:rq4}, as the LLM must resolve or ignore such contradictions when generating code.

\vspace*{-0.5\baselineskip}
\subsection{Adaptive Mutation Algorithm}
\label{sec:algorithm}

The mutation process, shown in \autoref{alg:reproducer}, produces $N$ structurally distinct mutated grammars from the original grammar. It employs an adaptive strategy to maintain diversity as the mutation space is explored.

Each iteration begins from the \emph{original} grammar tree and applies $m$ independently and uniformly selected mutation operators in sequence. The result is hashed and compared against all previously generated mutants. If the mutant is a duplicate, a repetition counter $r$ is incremented. After 5 consecutive duplicates, the mutation count $m$ is increased by one, encouraging more aggressive mutations to escape saturated regions of the space. If the mutant is novel, it proceeds to LLM synthesis.

Because each operator is drawn uniformly at random at each step, every operator has an equal probability of selection. This property is important for the mutation effectiveness analysis in Section~\ref{sec:rq3}: the number of findings attributable to each operator reflects its natural productivity under uniform sampling, without requiring separate per-operator experimental runs.

\begin{algorithm2e}[t]
	\small
	\SetAlgoLined
	\SetAlgoNoEnd
	\SetAlgoInsideSkip{-1pt}
	\SetAlCapSkip{2pt}
	\setlength{\algomargin}{3pt}
	\SetInd{0.3em}{0.5em}
	\caption{\small Adaptive Grammar Mutation and Input Generation}
	\label{alg:reproducer}
	\Input{Grammar $G$; Target mutant count $N$; Inputs per grammar $K$}
	\Output{Input set $S$}
	$S \assignment \emptyset$; $\mathcal{H} \assignment \emptyset$; $m \assignment 1$; $r \assignment 0$\ \Comment*[r]{r is repetition counter}
	$T \assignment \textsc{ParseGrammar}(G)$; $\mathcal{H} \assignment \mathcal{H} \cup \{\textsc{Hash}(T)\}$\;
	\While{$|\mathcal{H}| < N$}{
		$T' \assignment T$\ \Comment*[r]{start from original}
		\For{$i \assignment 1$ \KwTo $m$}{$T' \assignment \textsc{Mutate}(T')$\;}
		$h \assignment \textsc{Hash}(T')$\;
		\If{$h \in \mathcal{H}$}{
			$r \assignment r + 1$\;
			\lIf{$r = 5$}{$r \assignment 0$; $m \assignment m + 1$}
			\Continue\;
		}
		$\mathcal{H} \assignment \mathcal{H} \cup \{h\}$; $r \assignment 0$\;
		$\textit{script} \assignment \textsc{LLMGenerate}(T')$\;
		\lIf{$\textsc{HasError}(\textit{script})$}{$\textit{script} \assignment \textsc{LLMRepair}(\textit{script},\, \textsc{GetError}(\textit{script}))$}
		$S \assignment S \cup \textsc{Execute}(\textit{script},\, K)$\;
	}
	\Return{$S$}
\end{algorithm2e}

\vspace*{-0.5\baselineskip}
\subsection{LLM-Based Code Synthesis}
\label{sec:llm-synthesis}

Each mutated grammar tree is serialized into a prompt and sent to a large language model with the instruction to produce a standalone Python script that generates binary test cases conforming to the mutated specification. The prompt provides all production rules and constraints and explicitly instructs the LLM not to modify the grammar, to ignore unsatisfiable constraints, and to include fallback generation to guarantee termination and output. Full prompt templates are provided in ~\ref{app:prompts}.

If the generated script fails to execute, the error traceback is fed back to the LLM in a bounded repair loop. Once a valid script is obtained, the driver executes it to produce $K$ test cases per grammar. The specific model and parameter values are reported in Section~\ref{sec:setup}. 

\section{Evaluation}
\label{sec:evaluation}
We evaluate the pipeline on PKCS\#1~v1.5 signature structure, addressing four research questions:

\begin{itemize}
	\item \textbf{RQ1.} Can the pipeline reproduce known specification violations identified by Morpheus?
	\item \textbf{RQ2.} Does the pipeline find discrepancies beyond those reported by Morpheus?
	\item \textbf{RQ3.} Which categories of grammar mutations are most effective at producing specification-violating inputs?
	\item \textbf{RQ4.} How reliably does the LLM generate code that faithfully implements mutated grammars?
\end{itemize}

\subsection{Experimental Setup}
\label{sec:setup}

We generated $N = 1{,}500$ mutated grammars from the PKCS\#1~v1.5 grammar described in Section~\ref{sec:grammar}, producing $K = 20$ test cases per grammar for a total of 30{,}000 test cases. Each test case was executed against all target library harnesses. The value of $N = 1{,}500$ was determined by a 24-hour time budget for the mutation and generation pipeline.

\textbf{Targets.} We test against 48 cryptographic library implementations. Of these, the majority are drawn from the original Morpheus evaluation~\cite{yahyazadeh2021morpheus}, plus 3 targets not previously tested by Morpheus: wolfSSL~v5.9.1, mbedTLS~v3.6.4, and OpenSSL~v3.5.3.

\textbf{LLM model.} We use OpenAI GPT-5-nano as the primary model.

\textbf{Baseline.} We use the formally proven oracle from Morpheus~\cite{yahyazadeh2021morpheus} as ground truth. Morpheus identifies 13 distinct non-trivial violation categories across the tested implementations: 5 signature forgery (SF), 7 minor leniency (ML), and 1 buffer overflow (BO). For each known violation, we check whether any test case generated by our pipeline triggers the same differential behavior: the vulnerable library accepts the input while the oracle rejects it.

\subsection{RQ1: Reproduction of Known Violations}
\label{sec:rq1}

Table~\ref{tab:rq1} summarizes the results. Out of 13 non-trivial violation categories reported by Morpheus, our pipeline reproduces 10 (76.9\% reproduction rate). All 5 signature forgery categories--the most security-critical violations--are successfully reproduced.

\begin{table}[t]
	\centering
	\caption{Reproduction of Morpheus's known specification violations. SF - Signature Forgery, ML - Minor Leniency, BO - Buffer Overflow.}
	\label{tab:rq1}
	\footnotesize
	\setlength{\tabcolsep}{3pt}
	\begin{tabular}{clc}
		\toprule
		\textbf{Bug} & \textbf{Description} & \textbf{Repr.} \\
		\midrule
		SF\#1 & Lax algorithm ID \& padding length & \cmark \\
		SF\#2 & No hash parameter check & \cmark \\
		SF\#3 & Accepting trailing bytes & \cmark \\
		SF\#4 & Lax prefix byte check & \cmark \\
		SF\#5 & Lax padding check & \cmark \\
		\midrule
		ML\#1 & Lax DER length octet & \cmark \\
		ML\#2 & Lax identifier octet form bit & \xmark \\
		ML\#3 & Incorrect extended tag decoding & \cmark \\
		ML\#4 & Lax DigestInfo length octet & \cmark \\
		ML\#5 & Lax hash function ID octets & \xmark \\
		ML\#6 & Lax prefix byte check & \cmark \\
		ML\#7 & Accepting $<$ 8 bytes of padding & \xmark \\
		\midrule
		BO\#1 & Buffer overflow on malformed input & \cmark \\
		\midrule
		\multicolumn{2}{l}{\textbf{Total: 10 / 13 reproduced}} & \textbf{76.9\%} \\
		\bottomrule
	\end{tabular}
\end{table}

\textbf{Analysis of reproduced violations.} The pipeline is most effective at reproducing violations that correspond to single-field structural deviations. SF\#3 (trailing garbage) is triggered by a single operator (\texttt{add\_garbage}), while SF\#4 (lax prefix check) is triggered by 8 different operators, indicating that many structural mutations resulted in inputs with modified prefix bytes. Notably, all 5 signature forgery categories are reproduced--these are the most security-critical violations, as they can enable signature forgery attacks.

\textbf{Analysis of missed violations.} The 3 missed violations (ML\#2, ML\#5, ML\#7) were not missed due to limitations in the mutation operators. In each case, the mutator produced grammars that, if faithfully implemented, should have triggered the corresponding violation. The misses are attributable to LLM hallucination (Section~\ref{sec:rq4}): the LLM-generated scripts failed to faithfully implement the mutated grammar. For example, ML\#2 (leniency in checking form bit of an identifier octet) requires a modification to a single type field and the rest of the structure left untouched--but LLM either reverts to standard-compliant type value (H2) or computes the padding length incorrectly (H1), producing inputs that do not test the intended violation.

\subsection{RQ2: New Discrepancies}
\label{sec:rq2}

We examined the differential behavior matrix for accept/reject disagreements not corresponding to any known Morpheus violation.

\textbf{Finding.} We discovered a previously unreported discrepancy in LibTomCrypt~v1.18.2. The library accepted an input in which the leading \texttt{0x00} byte of the PKCS\#1~v1.5 structure was absent, while the remainder of the structure--block type, padding, separator, and DigestInfo--was well-formed. The missing leading byte was effectively absorbed into the padding region: the malformed input begins directly with block type \texttt{0x01} followed by the \texttt{0xFF} padding bytes, separator \texttt{0x00}, and a valid DigestInfo. The same behavior was observed in CryptX~v0.070, which is expected since CryptX uses LibTomCrypt as its underlying cryptographic library--the discrepancy originates in LibTomCrypt's PKCS\#1~v1.5 verification logic.

According to RFC~8017, the encoded message must begin with \texttt{0x00~||~0x01}. The absence of the leading zero byte constitutes a specification violation. All other tested libraries correctly reject this input. This discrepancy was triggered by the \texttt{remove\_component} mutation operator, which removed the \texttt{<leading\_byte>} nonterminal from the grammar. This finding was not reported by Morpheus, demonstrating that grammar-level mutation--specifically, removal of structural components--can explore input regions that combinatorial approaches do not cover.

\textbf{New targets.} For the three newly added targets (wolfSSL~v5.9.1, mbedTLS~v3.6.4, OpenSSL~v3.5.3), both our pipeline and Morpheus found only incompatibility issues, indicating that these current-version implementations correctly handle the PKCS\#1~v1.5 verification path for the tested configurations.

\subsection{RQ3: Mutation Operator Effectiveness}
\label{sec:rq3}

Since each mutation operator is selected uniformly at random (Section~\ref{sec:algorithm}), we can attribute each finding to the specific operator(s) that produced the mutated grammar triggering it. Table~\ref{tab:rq3} reports the number of distinct violation categories triggered by each operator, alongside the total number of mutants in which each operator appeared.

\begin{table}[t]
	\centering
	\caption{Mutation operator effectiveness. ``Findings'' counts distinct violation categories triggered. ``Appearances'' counts mutants containing this operator.}
	\label{tab:rq3}
	\footnotesize
	\begin{tabular}{llcc}
		\toprule
		\textbf{Category} & \textbf{Operator} & \textbf{Appear.} & \textbf{Findings} \\
		\midrule
		\multirow{9}{*}{\shortstack[l]{Structural\\(TLV-ign.)}}
		& \texttt{swap\_value}         & 246 & 4 \\
		& \texttt{reuse\_value}        & 234 & 4 \\
		& \texttt{swap\_order}         & 212 & 4 \\
		& \texttt{add\_garbage}        & 221 & 3 \\
		& \texttt{repeat\_value}       & 231 & 3 \\
		& \texttt{long\_length\_form}  & 215 & 3 \\
		& \texttt{byte\_steal}         & 197 & 3 \\
		& \texttt{extend\_tag}         & 226 & 2 \\
		& \texttt{remove\_component}   & 217 & 2 \\
		\midrule
		Struct.\ (TLV-aw.) & \texttt{change\_len\_val} & 169 & 3 \\
		\midrule
		\multirow{3}{*}{Constraint}
		& \texttt{remove\_constr}      & 191 & 0 \\
		& \texttt{flip\_signs}         & 216 & 0 \\
		& \texttt{exchange\_subexpr}   & 235 & 0 \\
		\bottomrule
	\end{tabular}
\end{table}

\textbf{Observations.} TLV-ignorant mutations are the most productive category, with \texttt{swap\_value}, \texttt{reuse\_value}, and \texttt{swap\_order} each triggering 4 distinct violation categories. The TLV-aware operator \texttt{change\_length\_change\_value} triggers 3 categories despite appearing in fewer mutants (169) than most TLV-ignorant operators, suggesting that preserving length-value consistency while modifying content is an efficient strategy for triggering leniency bugs.

Notably, \textbf{none of the three constraint mutation operators produced any findings}, despite appearing in a substantial number of mutants (191-235 each). This is not because constraint mutations are inherently ineffective--rather, it reflects the interaction between constraint mutations and LLM hallucination, as analyzed in Section~\ref{sec:rq4}. The operators appear frequently enough that they had ample opportunity to trigger violations; the bottleneck is the LLM's inability to faithfully implement the mutated constraints.

\textbf{Time budget justification.} The 1{,}500 mutated grammars were generated within a 24-hour time budget. The operators are distributed approximately uniformly across the corpus (169-246 appearances per operator), indicating that all operators had sufficient representation. The reproduction of 10 out of 13 violation categories, including all 5 security-critical signature forgery categories, suggests that the corpus provides adequate coverage of the mutation space.
                  
\subsection{RQ4: LLM Hallucination Analysis}
\label{sec:rq4}

This section presents the central analytical contribution of this work: a quantitative characterization of how LLMs hallucinate when generating code from formal grammars with semantic constraints.

\subsubsection{Operational vs.\ Semantic Reliability}

Of the 1{,}500 mutated grammars, GPT-5-nano produced scripts that executed successfully on the first attempt for 1{,}442 (96.1\%). An additional 55 (3.7\%) required the repair loop and were successfully repaired. Only 3 mutants (0.2\%) resulted in total generation failure. Operationally, the LLM achieves a 99.8\% success rate.

However, execution success does not imply semantic correctness. We classified all 1{,}500 generated scripts by whether they faithfully implement the mutated grammar. Of the 1{,}500 scripts, only 262 (17.5\%) correctly implement the mutated specification. The remaining 1{,}238 (82.5\%) exhibit some form of hallucination--the generated code deviates from the provided grammar in semantically significant ways. This reveals a striking gap between operational reliability (99.8\%) and semantic fidelity (17.5\%).

\subsubsection{Hallucination Taxonomy}

We identify five distinct hallucination types, distinguished by whether the LLM followed the intended mutation and the nature of the deviation. Table~\ref{tab:hallucination} presents the full breakdown.

\begin{table}[t]
	\centering
	\caption{LLM hallucination taxonomy across all 1{,}500 generated scripts.}
	\label{tab:hallucination}
	\footnotesize
	\begin{tabular}{clccc}
		\toprule
		\textbf{Type} & \textbf{Description} & \textbf{Mut.\ followed?} & \textbf{Count} & \textbf{\%} \\
		\midrule
		H1 & Padding miscalculation       & Yes     & 402 & 26.8 \\
		H2 & Mutation fully ignored        & No      & 17  & 1.1 \\
		H3 & Partial (contradictory grammar) & Partial & 440 & 29.3 \\
		H4 & Partial (no apparent reason)  & Partial & 270 & 18.0 \\
		H5 & Degenerate output             & N/A     & 106 & 7.1 \\
		\midrule
		& Correct                       & Yes     & 262 & 17.5 \\
		& Failed (no script produced)   & N/A     & 3   & 0.2 \\
		\midrule
		& \textbf{Total}                &         & \textbf{1{,}500} & \textbf{100} \\
		\bottomrule
	\end{tabular}
\end{table}

\textbf{H1 (padding miscalculation, 26.8\%)} scripts correctly implement the intended mutation but compute the padding length incorrectly --using a hardcoded value or a simplified formula that does not account for the mutation's effect on other field sizes. The padding length in PKCS\#1~v1.5 must be derived as the difference between the total message length and the sum of all other component lengths--a multi-step arithmetic derivation that the LLM frequently simplifies. These scripts are \emph{partial successes}: the mutation is tested, but the message structure is malformed in an unintended way. Depending on the target's parsing behavior, H1 scripts may still trigger the intended violation or may be rejected for the wrong reason.

\textbf{H2 (mutation fully ignored, 1.1\%)} scripts produce standard-compliant PKCS\#1~v1.5 structures despite receiving a mutated grammar. The LLM appears to override the provided specification with its prior knowledge of what a valid PKCS\#1 message looks like. Although rare in absolute terms, H2 is the most consequential hallucination type per occurrence: the mutation is not tested at all. The prompt explicitly instructs the LLM not to modify or fix the grammar (Section~\ref{sec:llm-synthesis}), yet these scripts disregard that instruction entirely.

\textbf{H3 (partially ignored, contradictory grammar, 29.3\%)} is the most prevalent hallucination type. These scripts partially implement the mutation but drop some constraints because the mutated grammar contains contradictions (see Section~\ref{sec:mutations}). For example, when \texttt{flip\_signs} changes the padding minimum from $\geq 16$ to $< 16$, this may conflict with the total length constraint. The LLM cannot satisfy both and silently drops one. H3 is an \emph{expected} limitation: a human would also need to make a choice when faced with contradictory requirements. This finding suggests that mutation operators should be designed to avoid producing contradictory grammars, or that a constraint satisfiability check should precede LLM synthesis.

\textbf{H4 (partially ignored, no apparent reason, 18.0\%)} scripts partially implement the mutation despite the mutated grammar being satisfiable. The LLM implements some aspects of the mutation but ignores others without justification. Unlike H3, this represents genuine unreliability: the grammar is correct and implementable, but the LLM fails to follow it fully. H4 is the most concerning hallucination type from a reliability standpoint, as it cannot be predicted or prevented through better mutation design.

\textbf{H5 (degenerate output, 7.1\%)} scripts produce random strings with no structural resemblance to any grammar--the LLM's fallback behavior when it cannot parse the specification.

\textbf{Output variation.} We also measured whether each script produces test cases of varying length (as opposed to identical outputs). Across all hallucination types, variation rates are consistently high: 100\% for H1, H2, and H5, 98.9\% for H3, and 99.6\% for H4. This indicates that hallucinated scripts still exhibit surface-level diversity--they appear to be functioning correctly by producing varied outputs--making hallucination difficult to detect without semantic analysis of the generated code.

\subsubsection{Hallucination by Mutation Category}

To determine whether hallucination patterns vary systematically across mutation types, we cross-tabulate hallucination type against the mutation category of each mutant. Since mutants produced by the adaptive algorithm may involve multiple operators (when $m > 1$), we classify each mutant by the combination of categories present in its operator list. Table~\ref{tab:cross} presents the results for the three pure categories (single-category mutants) and the three mixed categories.

\begin{table*}[t]
	\centering
	\caption{Hallucination type by mutation category (\% within each category). TLV-ign -- structural TLV-ignorant, TLV-aw -- structural TLV-aware, Constr -- constraint.}
	\label{tab:cross}
	\footnotesize
	\begin{tabular}{lcccccc}
		\toprule
		& \textbf{TLV-ign.} & \textbf{TLV-aw.} & \textbf{Constr.} & \textbf{TLV-ign.+aw.} & \textbf{TLV-ign.+constr.} & \textbf{TLV-aw.+constr.} \\
		& ($n = 796$) & ($n = 17$) & ($n = 99$) & ($n = 104$) & ($n = 441$) & ($n = 43$) \\
		\midrule
		Correct & 13.3\% & 17.6\% & 30.3\% & 4.8\% & 26.1\% & 7.0\% \\
		H1      & 23.9\% & 35.3\% & 33.3\% & 25.0\% & 30.2\% & 32.6\% \\
		H2      & 0.8\%  & 0.0\%  & 8.1\%  & 1.0\%  & 0.5\%  & 0.0\% \\
		H3      & 32.4\% & 29.4\% & 13.1\% & 39.4\% & 23.8\% & 41.9\% \\
		H4      & 21.5\% & 11.8\% & 9.1\%  & 19.2\% & 13.6\% & 18.6\% \\
		H5      & 8.2\%  & 5.9\%  & 5.1\%  & 10.6\% & 5.4\%  & 0.0\% \\
		\bottomrule
	\end{tabular}
\end{table*}

\noindent\textbf{Key findings from the cross-tabulation:}

\emph{Constraint-only mutations have the highest correct rate (30.3\%) but also the highest H2 rate (8.1\%).} This is a striking pattern: when the grammar's production rules remain unchanged and only a constraint is mutated, the LLM is more likely to produce correct output--but is also most likely to ignore the mutation entirely. The structural similarity to a standard PKCS\#1 message may make it easier for the LLM to ``recognize'' the format and generate correct code, but the same recognition may cause it to override the mutated constraint with its prior knowledge.

\emph{Pure TLV-ignorant mutations have the highest H3 rate (32.4\%) and H4 rate (21.5\%).} When only structural operators are applied, the resulting grammar frequently contains implicit contradictions (e.g., swapping two field values creates a length mismatch) that the LLM cannot resolve (H3), or produces grammars that the LLM partially ignores for no identifiable reason (H4).

\emph{Mixed categories show compounding effects.} Mutants combining TLV-ignorant and TLV-aware operators have the lowest correct rate (4.8\%) and highest H5 rate (10.6\%), suggesting that the combination of structural changes at different levels of TLV awareness produces grammars that are particularly difficult for the LLM to interpret.

\emph{H1 (padding miscalculation) is prevalent across all categories (23.9-35.3\%),} confirming that incorrect padding computation is a general LLM weakness independent of mutation type. This is consistent with the finding that padding length derivation requires multi-step arithmetic--a known difficulty for LLMs.

\textbf{Implications for constraint mutations and RQ3.} The cross-tabulation explains why constraint mutations produced zero findings in RQ3 despite adequate representation. Constraint-only mutants achieve 30.3\% correctness and 33.3\% H1 (which partially tests the mutation), but 8.1\% are fully ignored (H2) and 13.1\% are partially ignored due to contradictions (H3). More importantly, constraint mutations appear predominantly in mixed-category mutants (441 TLV-ignorant+constraint vs.\ 99 pure constraint), where the combined mutations create additional complexity that reduces correctness to 26.1\%. The net effect is that constraint mutations rarely produce test cases that accurately reflect the intended constraint violation.

\subsubsection{Impact on Bug Detection}

To close the loop between hallucination and bug detection, we trace each of the 3 missed violation categories to the hallucination behavior that prevented detection:

\begin{itemize}
	\item \textbf{ML\#2} (lax identifier octet form bit check): requires inputs with modified identifier octets. The mutator produced relevant grammars, but the LLM either reverted to standard identifier values (H2) or computed padding incorrectly (H1), preventing the specific field modification from reaching the target.
	\item \textbf{ML\#5} (lax hash function ID content octets): requires precise modification of the OID bytes within the DigestInfo. The LLM's H4 behavior (partial implementation) resulted in scripts that modified some fields but left the OID unchanged.
	\item \textbf{ML\#7} (accepting $<$ 8 bytes of padding): directly requires a constraint mutation (\texttt{flip\_signs} on the padding minimum). As discussed above, the LLM either ignores the flipped constraint (H2) or cannot reconcile it with the total length constraint (H3).
\end{itemize}

\noindent This traceability is itself a property of the approach: every missed violation has a concrete, identifiable cause rooted in a specific hallucination type--not merely ``insufficient exploration.'' This level of diagnostic transparency is unusual in automated test generation and directly relevant to the goal of building explainable testing tools. 	\section{Discussion}
\label{sec:discussion}

\subsection{Key Takeaways}

The 76.9\% reproduction rate--including all 5 signature forgery categories --demonstrates that grammar-level mutation combined with LLM code synthesis is a viable approach to specification compliance testing without domain-specific engineering. The new discrepancy in LibTomCrypt confirms that grammar-level operations (removing a structural component) can reach input regions that combinatorial approaches miss. However, the 3 missed violations are entirely attributable to LLM hallucination, not the mutation strategy, suggesting the approach's ceiling is higher than current results indicate.

The most significant finding is the gap between operational reliability (99.8\%) and semantic fidelity (17.5\%). This gap is masked by high output variation rates (98.9-100\% across all hallucination types), making hallucination difficult to detect without semantic analysis. Any approach that delegates code generation to an LLM based on a formal specification must contend with this gap--operational metrics alone are insufficient.

The hallucination taxonomy yields actionable guidance: H3 (29.3\%, contradictory grammars) is a mutation design problem addressable by pre-synthesis satisfiability checks; H4 (18.0\%, unexplained partial deviation) is an irreducible LLM reliability problem requiring post-generation verification or more capable models; H1 (26.8\%, padding miscalculation) reflects a general arithmetic weakness uniform across mutation categories; and H2 (1.1\%, mutation fully ignored) is rare but the most consequential per occurrence.

\subsection{Limitations}

\textbf{Single protocol.} The evaluation targets only PKCS\#1~v1.5. While the pipeline is protocol-agnostic in design, its effectiveness on protocols with different structural characteristics (e.g., text-based protocols, stateful protocols, protocols without strict length-value constraints) has not been evaluated.

\noindent\textbf{Manual grammar construction.} The grammar was manually derived from the RFC. Errors in the grammar could produce false positives or false negatives. Automating grammar extraction from RFC documents is an important direction for future work.

\noindent\textbf{Single LLM model.} We evaluated a single LLM. Different models may exhibit different hallucination rates and type distributions. Whether the patterns we observe -- particularly the dominance of H3 and the prevalence of H1 across categories -- generalize across models remains an open question.

\noindent\textbf{Hallucination classification.} The classification of scripts into hallucination types involves judgment calls, particularly at the boundary between H3 (contradictory grammar) and H4 (no apparent reason). We mitigated this by defining H3 based on whether the mutated grammar's constraints are simultaneously satisfiable, but edge cases exist where satisfiability is ambiguous.

\subsection{Future Work}

\textbf{Post-generation verification.} The most direct improvement would be to verify generated test cases against the mutated grammar's constraints before including them in the corpus. A lightweight checker that validates TLV structure and length-field consistency could filter out H1, H2, and H5 hallucinations automatically, and flag H3/H4 cases for closer inspection.

\noindent\textbf{Contradiction-free mutation.} Since H3 accounts for 29.3\% of all scripts, designing mutation operators that guarantee constraint satisfiability -- or adding a satisfiability pre-check -- could substantially improve effective fidelity.

\noindent\textbf{Extension to additional protocols.} Evaluating the approach on other TLV-encoded protocols (X.509 certificates, CMS/PKCS\#7, OCSP) would test generalizability and determine whether the hallucination patterns observed for PKCS\#1~v1.5 are protocol-specific or fundamental.

\noindent\textbf{Automated grammar extraction.} Recent work has demonstrated LLM-based grammar extraction for text-based protocols~\cite{liu2025synthesizing}. Extending this to binary protocols with ASN.1/DER specifications would eliminate the manual grammar construction step. 	

\section{Related Work}
\label{sec:related}

\textbf{Grammar-based testing and grammar mutation.}
Gmutator~\cite{bendrissou2025grammar} is the most closely related work: it mutates ANTLR production rules to create mutant grammars for parser testing, but operates only on syntactic structure and explicitly identifies constraint mutation as future work.
Our approach extends Gmutator by adding constraint mutations, targeting binary TLV-encoded protocols, and using LLM-based code synthesis instead of direct grammar-based generation. FANDANGO~\cite{zamudio2025fandango} evolves inputs within a fixed grammar using a genetic algorithm; unlike our approach, it does not mutate the grammar itself. FuzzEval~\cite{hasan2024fuzzeval} showed that eleven state-of-the-art fuzzers fail to produce valid context-sensitive inputs for PKCS\#1~v1.5, motivating our grammar-guided approach.

\noindent\textbf{LLM-assisted testing.}
ChatAFL~\cite{meng2024chatafl} uses an LLM to extract protocol grammars and enrich seeds for text-based protocols; the LLM's role is grammar extraction, whereas ours is code synthesis. SeedMind~\cite{shi2024harnessing} uses LLMs to generate input generators from harness code; our approach is specification-driven rather than implementation-driven. Neither evaluates the fidelity of LLM-generated artifacts--the central question of our work.

\noindent\textbf{Protocol compliance testing.}
Morpheus~\cite{yahyazadeh2021morpheus}, our primary baseline, constructs a formally verified combinatorial model for PKCS\#1~v1.5 and achieves exhaustive coverage but requires deep protocol expertise. SAECRED~\cite{dar2025saecred} and AFLNET~\cite{pham2020aflnet} target stateful protocols; SAECRED is domain-specific (SAE/WPA3), while AFLNET operates without a specification and cannot detect compliance violations.

\noindent\textbf{LLM reliability in code generation.}
Liu et al.~\cite{liu2026beyond} identified three primary categories of code hallucination (Requirement Conflicting, Code Inconsistency, Knowledge) with 12 specific types; Zhang et al.~\cite{zhang2025llm} showed that hallucination rates increase with contextual complexity. Both focus on general-purpose code generation from natural language. Our taxonomy addresses the unstudied setting of generating code from formal specifications with arithmetic constraints, and introduces the distinction between hallucinations with identifiable causes (H3: contradictory grammar) and those without (H4: unexplained deviation).

\section{Conclusion}
\label{sec:conclusion}

We investigated whether grammar-level mutation combined with LLM-based code synthesis is a viable approach to specification compliance testing of binary protocols. Evaluating on PKCS\#1~v1.5 across 48 implementations, we found that the approach reproduces 10 of 13 known violation categories -- including all 5 signature forgery categories -- and uncovers 1 previously unreported discrepancy, without requiring domain-specific engineering.

Our central finding is that LLM semantic fidelity, not the mutation strategy, is the primary bottleneck: 82.5\% of generated scripts deviate from the provided grammar despite 99.8\% executing successfully. We identified five hallucination types and showed that their distribution varies systematically by mutation category, yielding actionable guidance: contradictory grammars (H3, 29.3\%) can be filtered by pre-synthesis satisfiability checks, while unexplained deviations (H4, 18.0\%) require post-generation verification. These findings apply broadly to any approach that uses LLMs to generate code from formal specifications.

\bibliographystyle{ACM-Reference-Format}
	\balance
	\bibliography{ref}
	\clearpage
	\appendix

\section{Prompt Templates}
\label{app:prompts}

This appendix provides the full prompt templates used for LLM-based code synthesis. Placeholders in curly braces (\texttt{\{grammar\}}, \texttt{\{inputs\}}, \texttt{\{seed\}}, \texttt{\{error\_message\}}) are substituted at runtime with the serialized mutated grammar and execution parameters.

\subsection{Generation Prompt}

The generation prompt consists of a system message that defines the LLM's role and output contract, and a user message that provides the mutated grammar.

\smallskip
\noindent\textbf{System message:}

\begin{quote}
	\small
	You are a professional system security engineer proficient in Python. Write a COMPLETE, runnable Python 3 script. The script generates testcases from the provided grammar.
	
	\textbf{Output requirements:} Output ONLY a single markdown code block containing the full script. No explanations, no prose outside the code block.
	
	\textbf{Script requirements:} Python 3; standard library only; no third-party modules; no classes (functions and module-level constants only); no assert statements; no sys.exit(); no uncaught exceptions; script MUST ALWAYS terminate. Must define: \texttt{generate(inputs: int) -> None}. Script MUST be executable as: \texttt{python3 generator.py inputs seed}. Script MUST call \texttt{random.seed(seed)} before generating testcases. Script MUST parse arguments using \texttt{sys.argv}.
	
	\textbf{Output format:} The script MUST print EXACTLY ONE LINE to STDOUT. That line MUST contain: exactly \texttt{\{inputs\}} testcases, separated by comma character `\texttt{,}', NO spaces anywhere, NO trailing comma, NO extra newline beyond standard print newline.
	
	\textbf{Failure handling:} The script MUST ALWAYS print valid output even if grammar parsing fails, generation fails, or an unexpected error occurs. In failure case: generate fallback testcases using simple random strings.
	
	\textbf{Grammar rules:} Do NOT modify the grammar text. Do NOT attempt to fix the grammar text. If constraints cannot be enforced, ignore them. Use best-effort generation. Prefer valid outputs when possible.
	
	\textbf{Diversity requirements:} Generation MUST use randomness to produce diverse outputs by varying: rule choices, recursion depth, expansion length, boundary values.
	
	\textbf{Safety requirements:} To guarantee termination: limit recursion depth, limit expansion steps, include fallback generation. Return ONLY the code block.
\end{quote}

\noindent\textbf{User message:}

\begin{quote}
	\small
	Write generator.py according to the specification.
	
	Grammar: \texttt{\{grammar\}}
	
	Inputs: \texttt{\{inputs\}}
	
	Seed: \texttt{\{seed\}}
\end{quote}

\subsection{Repair Prompt}

The repair prompt is sent when a generated script fails to execute. It provides the error traceback and reiterates the full output contract.

\smallskip
\noindent\textbf{User message:}

\begin{quote}
	\small
	You previously produced a Python script named generator.py to generate testcases from a grammar. The script failed. Your task: Produce a FULL corrected Python 3 script. Output ONLY the complete corrected script inside a single markdown fenced code block. Do NOT include any explanations.
	
	\textbf{Script contract (all must hold after fix):} Python 3; standard library only; no third-party modules; no classes; no assert statements; no sys.exit(); no uncaught exceptions; script MUST always terminate. Must define: \texttt{generate(inputs: int) -> None}. Script MUST work exactly like: \texttt{python3 generator.py inputs seed}. Parse inputs from \texttt{sys.argv[1]}, parse seed from \texttt{sys.argv[2]}, call \texttt{random.seed(seed)} BEFORE generating testcases.
	
	\textbf{Output contract:} Script MUST print EXACTLY ONE LINE to STDOUT containing exactly \texttt{\{inputs\}} testcases, separated by comma `\texttt{,}', NO spaces, NO trailing comma.
	
	\textbf{Failure handling:} Script MUST ALWAYS print valid output even if grammar parsing fails, generation fails, or an unexpected exception occurs. In failure case: generate fallback random testcases, still print exactly \texttt{\{inputs\}} testcases.
	
	\textbf{Required entrypoint:}
	\begin{verbatim}
		if __name__ == "__main__":
		import sys
		inputs = int(sys.argv[1])
		seed = int(sys.argv[2])
		random.seed(seed)
		generate(inputs)
	\end{verbatim}
	
	\textbf{Grammar rules:} Do NOT modify grammar text. Do NOT fix grammar text. Ignore unsatisfiable constraints. Use best-effort generation.
	
	\textbf{Fixing requirements:} Use the error traceback to identify root cause. Fix ALL issues including: crashes, wrong output format, wrong CLI parsing, wrong seed usage, infinite recursion, missing fallback. You may refactor code if necessary but preserve functionality when possible.
	
	\textbf{Termination requirement:} Script MUST always terminate. Use recursion limits, iteration limits, or fallback if needed.
	
	Error message / traceback: \texttt{\{error\_message\}}
	
	Inputs: \texttt{\{inputs\}}
	
	Seed: \texttt{\{seed\}}
\end{quote} \end{document}